\documentclass[useAMS,usenatbib,a4paper]{mn2e}

\usepackage[left=2cm,top=2cm,right=2cm,bottom=2cm]{geometry}
\usepackage{amssymb,epsfig,color}
\usepackage{amsmath}
\usepackage{graphicx}
\usepackage[caption=false]{caption}
\usepackage[font=footnotesize]{subfig}
\usepackage{multirow}
\usepackage{xspace}
\usepackage{hyperref}

\title[A quasi-periodic oscillation in the blazar J1359+4011]{A quasi-periodic oscillation in the blazar J1359+4011}
\author[O.G. King et al.]{O.G. King$^{1}$\thanks{E-mail: ogk@astro.caltech.edu}, 
T. Hovatta$^1$,
W. Max-Moerbeck$^2$, 
D.L. Meier$^3$,
T.J. Pearson$^1$, 
\newauthor 
A.C.S. Readhead$^1$,
R. Reeves$^1$, 
J.L. Richards$^4$,
M.C. Shepherd$^1$\\
$^{1}$California Institute of Technology, 1200 E California Blvd, MC 249-17, Pasadena, CA 91125, USA\\
$^{2}$National Radio Astronomy Observatory, P.O. Box 0, Socorro, NM 87801, USA\\
$^{3}$NASA Jet Propulsion Laboratory, 4800 Oak Grove Dr., Pasadena, CA 91109, USA\\
$^{4}$Department of Physics, Purdue University, West Lafayette, IN 47907, USA\\
}

\begin{document}

\maketitle

\begin{abstract}
The OVRO $40\,$m telescope has been monitoring the 15$\,$GHz radio flux density of over 1200 blazars since 2008. The 15$\,$GHz light curve of the flat spectrum radio quasar J1359+4011 shows a strong and persistent quasi-periodic oscillation. The time-scale of the oscillation varies between 120 and 150 days over a $\sim4$ year time span. We interpret this as the active galactic nucleus mass-scaled analog of low-frequency quasi-periodic oscillations from Galactic microquasars, or as evidence of modulation of the accretion flow by thermal instabilites in the ``inner'' accretion disc.
\end{abstract}


\section{Introduction}




Quasi-periodic behaviour in blazars at radio wavelengths occurs on time-scales of order years. It usually takes the form of periodic flaring, for instance in the BL~Lacertae object OJ~287 (1.12 or 1.66$\,$yr, \citealt{1998ApJ...503..662H}), BL~Lac itself ($\sim2$ and $\sim 8\,$yr, \citealt{2003MNRAS.341..405S,2004A&A...424..497V}), and AO~0235+16 ($\sim5$ or $\sim8\,$yr, \citealt{2001A&A...377..396R,2006A&A...459..731R}). This behaviour is usually explained by processes related to the orbital dynamics of the disc/jet system, such as periodically varying Doppler beaming from a precessing jet \citep{1985ApJ...298..114M,1992A&A...255...59C,2000A&A...355..915A,2013MNRAS.428..280C} or processes related to the innermost stable orbit of the accretion disc \citep{2006MNRAS.367..905B,2013arXiv1307.1113P}.

While quasi-periodic oscillations (QPOs) in the X-ray emission of stellar mass black-hole binaries are a common and well-studied phenomenon \citep{2006ARA&A..44...49R}, corresponding behaviour from supermassive black holes at the heart of AGN is very rare. Only one X-ray QPO from an AGN -- the narrow line Seyfert~1 galaxy REJ~1034+396 -- has been reliably measured \citep{2008Natur.455..369G,2012A&A...544A..80G}. It has a period of $\sim 1\,$h, and is thought to be the mass-scaled equivalent of high frequency QPOs seen in stellar black hole binaries \citep{2010MNRAS.403....9M}.

In this letter we describe the discovery of a remarkable and persistent quasi-periodic oscillation in the $15\,$GHz light curve of the flat-spectrum radio quasar CGRABS J1359+4011. The source  J1359+4011 ($13^{\rm h}59^{\rm m}38.1{\rm s}$, $+40^{\rm d}11^{\rm m}38.3{\rm s}$ J2000) is a high Galactic latitude ($b = 70.8^{\circ}$) flat-spectrum radio quasar (FSRQ) \citep{2001MNRAS.323..757L,2005ApJ...626...95S} at a redshift of $z = 0.407 \pm 0.001$. 
It has been detected at X-ray energies using \emph{ROSAT} \citep{2001MNRAS.323..757L}, but it is not visible at $\gamma$-ray energies and has not been included in either of the \emph{Fermi} catalogs \citep{2010ApJS..188..405A,2012ApJS..199...31N}. 
J1359+4011 is thought to contain a $10^8\,M_{\sun}$ black hole with high Eddington ratio accretion (Table~\ref{tab:properties_of_J1359+4011}).
The data are described in Section~\ref{sec:data}, the QPO is explored using a wavelet decomposition in Section~\ref{sec:wavelet}, and we interpret the QPO in Section~\ref{sec:interpretation} as being either the AGN analog of low-frequency QPOs, or as being caused by thermal instabilities in the ``inner'' accretion disc.

\section{Data} \label{sec:data}

The 40$\,$m telescope of the Owens Valley Radio Observatory (OVRO) is being used to monitor the 15$\,$GHz flux densities of over 1200 active galaxies twice a week. This program has been running continuously since 2008. The data are reduced and calibrated to form regularly updated light curves that are made available on the internet\footnote{\url{http://www.astro.caltech.edu/ovroblazars/}}.

The OVRO 40$\,$m receiver uses an alternation of two beams on the source \citep{1989ApJ...346..566R} to remove the varying atmospheric emission and diffuse background emission from the source flux density measurement. The data are calibrated and the flux density scale is set using the primary calibrator 3C286, as described in \citet{Richards:2011p5170}.

\begin{figure}
 \centering
 \includegraphics[width=9cm]{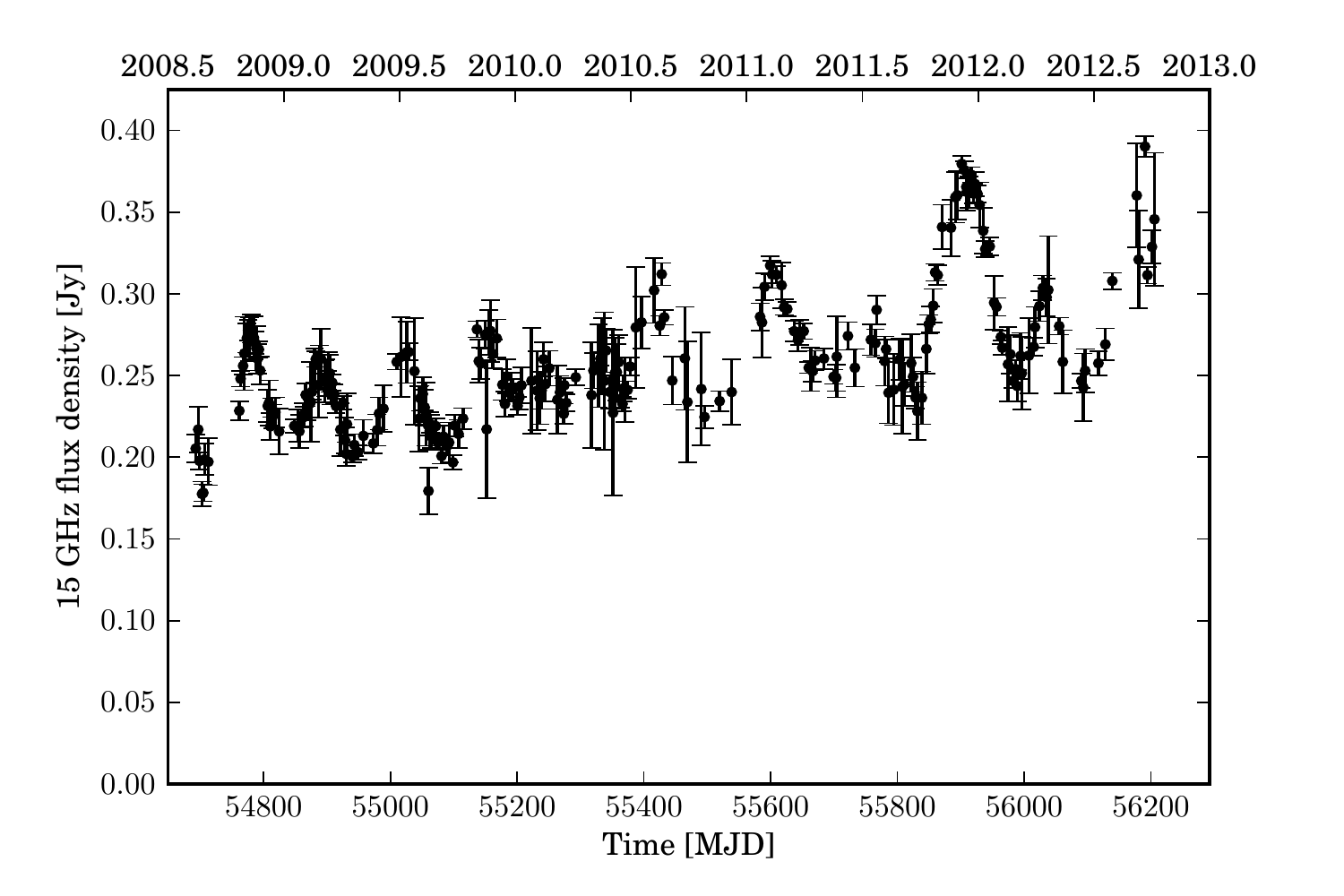}
 \caption{The OVRO 15$\,$GHz light curve for J1359+4011. The data cover a 4.1$\,$yr time span and show a strong quasi-periodic modulation (Fig.~\ref{fig:wavelet_wwz_with_light_curve}).}
 \label{fig:ovro_light_curve}
\end{figure}

The $4.1\,$yr light curve of J1359+4011 is shown in Fig.~\ref{fig:ovro_light_curve}. The object has been observed slightly more often than once a week, on average, with occasional gaps in the light curve due to telescope maintenance or poor weather.
A modulation of about 25\% of the flux density is visible. This is the only source in the OVRO catalogue to show such a strong and persistent modulation.

Two instrumental effects would produce such a modulation of the light curve. The first is related to the dual-beam measurement process. If the reference field around the target source were to contain a patch of bright emission, this might produce elevation dependent flux-density measurements. However, the data show no dependence of the flux density on elevation.
The second systematic effect that might cause the modulation is linked to the pointing procedure used by the OVRO 40$\,$m telescope. The sky is divided in to 134 regions and pointing corrections for the 40$\,$m telescope are calculated in each region. Time-dependent errors in the pointing correction for the region around J1359+4011 would result in a modulated light curve. 
Sources that share the same pointing correction would share the same fractional flux density modulation. However, careful study of the sources that share the same pointing correction shows that they contain no such common modulation. We conclude that the modulation seen in the J1359+4011 light curve is intrinsic to the source.

\section{Quasi-periodic behaviour} \label{sec:wavelet}


Wavelets are an ideal tool for measuring quasi-periodic fluctuations. The data are decomposed into \emph{localized} functions, which is preferable to Fourier analysis when searching for short-lived fluctuations or fluctuations with a varying period.
\citet{1996AJ....112.1709F} introduced an alternative to the discrete wavelet transform, the weighted wavelet Z-transform (WWZ), that is better suited to discovering the time-scale of fluctuations in the light curve and is robust against missing data. It is based on the Morlet wavelet \citep{Grossmann:1984} and has become a much-used tool in fields as diverse as paleoceanography \citep{2010PalOc..25.2218I}, climate change \citep{2010Ap&SS.326..181J}, pulsating variable stars \citep{2009ApJ...691.1470T}, and AGN variability \citep{2008A&A...488..897H}. The values that the WWZ takes can be thought of as  goodness-of-fit parameters similar to the familiar chi-square statistic.

\begin{figure}
 \centering
 \includegraphics[width=9cm]{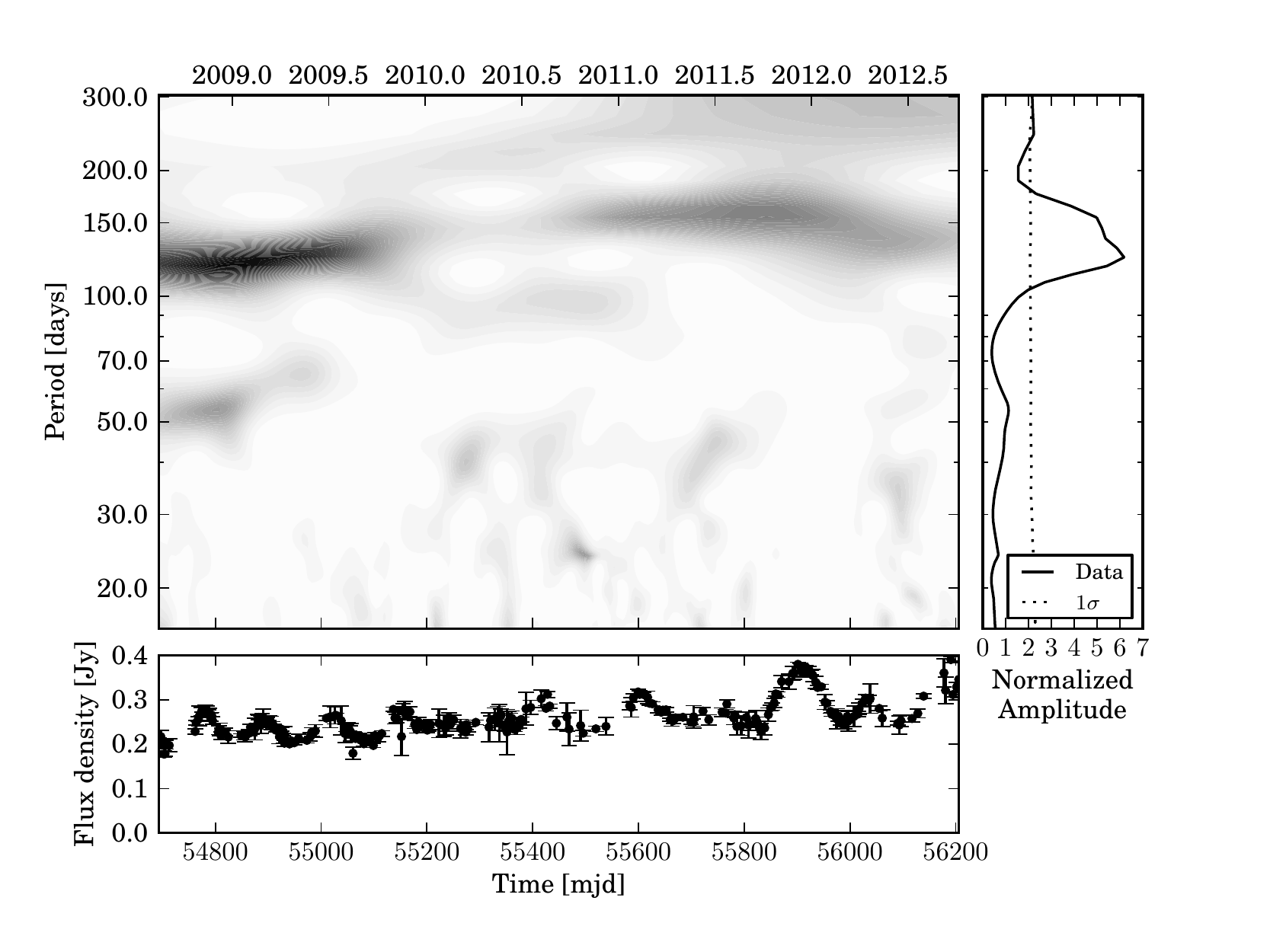}
\caption{The WWZ of the light curve for the source J1359+4011, with darker colour indicating a stronger signal. A strong signal is seen throughout 2008 and most of 2009. It appears to fade in strength and increase in duration before returning in late 2011. The WWZ amplitude has been normalized as described in the text; the global wavelet in the right panel includes the $1\sigma$ significance contour determined from Monte Carlo simulations.}
 \label{fig:wavelet_wwz_with_light_curve}
\end{figure}

We used Monte Carlo simulations to establish the statistical significance of the structures in the WWZ of the light curve. The null hypothesis is that the light curve is a Gaussian random process with power law power spectrum. We generated artificial light curves with the same PSD slope of $1.5$ as J1359+4011 using a modified implementation of \citet{2002MNRAS.332..231U}, as described in W. Max-Moerbeck et al. (in preparation). The artificial light curves had the same variance and sampling as the data. We used 1000 artificial light curves to establish the mean and standard deviation of the WWZ at each point in the period/time plane for the artificial data. The WWZ of J1359+4011 is shown in Fig.~\ref{fig:wavelet_wwz_with_light_curve}. The amplitude has been normalized by the mean established through Monte Carlo simulations. The average wavelet over time in the right panel includes the $1\sigma$ contour. 

The wavelet analysis reveals an initial $\sim 120\,$d oscillation in 2008/2009 that then fades in amplitude and increases in period, reaching a minimum amplitude in mid-2010. It then increases in amplitude and peaks at a period of $\sim150\,$d in early 2012. The range of detectable time-scales is from 15 to $300\,$d.

\begin{table}
\centering
\caption{The properties of J1359+4011 from \citet{2003MNRAS.339.1081D}. The mass was determined using the $H\beta$ emission line and the luminosity calculated in the interval $7.08\times 10^{14} - 7.24\times 10^{14}\,$Hz.
}
\label{tab:properties_of_J1359+4011}
\begin{tabular}{|c|c|c|c|c|}
\hline
$\log M_{\bullet}$ & $\log \dot{M}$ & $\log \left( \frac{L}{L_{\rm Edd}} \right)$ & $\log \left( \frac{L_{\rm th}}{L_{\rm tot}} \right)$ & $\log L_{\rm tot}$ \\ 
$\left( M_{\sun} \right)$ & $\left(M_{\sun}\,{\rm yr}^{-1}\right)$  &						&										  & $({\rm erg}\,{\rm s}^{-1}$ \\
 & & & &${\rm Hz}^{-1})$ \\ \hline
$8.0$ & $-1.7$ & $-1.6$ & $-1.7$ & $30.03$ \\ \hline
\end{tabular}
\end{table}

\section{Interpretation} \label{sec:interpretation}


A variety of mechanisms has been proposed to explain periodic behaviour in blazar radio and optical emission. These include periodically varying Doppler beaming from a precessing jet \citep{1985ApJ...298..114M,1992A&A...255...59C,2000A&A...355..915A,2013MNRAS.428..280C} and processes related to the innermost stable orbit of the accretion disc \citep{2006MNRAS.367..905B,2013arXiv1307.1113P}. The periods observed are generally on time-scales of order years.

One possible explanation for the behaviour of J1359+4011 is that the oscillation 
is a direct analog of the QPOs seen in microquasars (jet-producing black hole 
X-ray binaries). Since many black hole time-scales vary approximately 
inversely with black hole mass, scaling the oscillation in J1359+4011 in this 
manner to a $10\,M_{\sun}$ black hole would result in a QPO of a few hertz -- 
squarely in the range of a low-frequency QPO (LFQPO) for a typical microquasar. 
LFQPOs in the latter sources are usually classified as types A, B, or C \citep{2005ApJ...629..403C}. The closest analog of the oscillation in J1359+4011 may be the 
type A QPO, which is associated with high Eddington ratio accretion\footnote{Here 
we define ``high Eddington ratio'' as an accretion ratio $\dot{m}=\dot{M}/\dot{M}_{\rm Edd}$ at which, in standard thin 
disc theory, the disc ``inner'' region is expected to be radiation pressure dominated. 
For $M_{\bullet} = 10 M_{\sun}$, this is $\dot{m} \gtrsim 0.15$ and for 
$M_{\bullet} = 10^8 M_{\sun}$, a high Eddington ratio would be anything above 
$\dot{m} \gtrsim 0.02$.} 
and low $Q$ ($\nu / \Delta \nu$). Type A QPOs also have low fractional variability (only a 
few percent), but that measurement is performed in the X-ray; very high time 
resolution radio studies have not yet been done on QPO-producing microquasars so 
no comparable radio variability values exist yet. In principle, one might be able to 
verify the expected low variability by observing J1359+4011 in the X-ray, but given past 
problems in finding X-ray QPOs in AGN (not to mention the possible beamed jet
contamination of the X-ray flux in this blazar), such an observation would be 
very difficult. Furthermore, the currently popular model for classical microquasar 
QPOs (particularly types A and C) is Lense-Thirring precession of a geometrically 
thick, accretion torus near the central black hole \citep{2009MNRAS.397L.101I,2011MNRAS.415.2323I,2011MNRAS.418.2292M}. 
Such models even predict the X-ray flux / QPO frequency correlations seen in type C QPOs.  
If this torus precession would slightly precess the pointing direction of the J1359+4011 blazar's jet, this 
model may account for the radio oscillations seen in Fig.~\ref{fig:ovro_light_curve}.

Another potential explanation for the observed QPO is an instability in the disc/jet system.
Magnetically choked accretion flows (MCAFs) in general-relativistic 3D MHD simulations are seen to produce quasi-periodic fluctuations in the energy outflow efficiency of the relativistic jet close to the black hole \citep{2011MNRAS.418L..79T}. The dominant mode of these fluctuations has a period of $\sim 70\,r_g/c$ for a rapidly spinning black hole ($a=0.9375$), which is $\sim 1\,$d when the Schwarzschild radius $r_{g} \simeq 3\times 10^{11}\,$m for a $10^8M_{\sun}$ black hole. Slower-spinning black holes are expected to have longer periods \citep{2012MNRAS.423.3083M}. 
The different time-scale makes it improbable that MCAFs are the mechanism behind the QPO seen in the J1359+4011, assuming that it is even possible for the MCAF QPOs to radiatively transfer out to the optically thin radio region of the jet. So-called dynamo cycles \citep{1995ApJ...446..741B} in simulations of black hole accretion discs -- oscillations in the azimuthal magnetic field above the accretion disc -- have periods of $\sim 5000\mathrm{~s~}GM_{\bullet}/c^3$ \citep{2011ApJ...736..107O} which is $\sim 30\,$d for a $10^8M_{\sun}$ black hole. It is unclear, however, whether these oscillations can produce an observable signature in the radio jet.

A third possible model that would be applicable only to high Eddington ratio black hole 
systems involves an instability in the accretion flow that would alternately feed 
and starve the black hole with accreting material. This is the 
Lightman-Eardley secular (accretion flow) instability, in which the entire radiation 
pressure dominated ``inner'' region of the accretion disc behaves in a manner 
opposite to that normally expected: the accretion rate varies {\em inversely} 
with vertically integrated disc surface density \citep{1974ApJ...187L...1L}. The 
accretion flow then breaks up into rings of high surface density (which accrete 
slowly internally) and low surface density (which accrete rapidly internally). 
The time for the entire radiation pressure dominated disc region to empty (12 min 
$(M_{\bullet} / 10 \, M_{\sun})^{4/3} \, (\dot{m} / 0.3)^{2/3}$) is much too long 
(9,000 years for a $10^8 \, M_{\sun}$ black hole accreting at $\dot{m} \sim 0.025$;
\citealt{Meier:2012vk}) to explain the J1359+4011 oscillations. However, if the disc rings formed 
via thermal processes, then when they initially form at or near the outer edge of the 
``inner'' disc region each would accrete toward the black hole, separated roughly by the 
thermal time at or near that outer edge. The thermal time scale at the outer edge of the 
``inner'' disc region is $\sim$1s $(M_{\bullet} / 10 \, M_{\sun})^{8/7} \, (\dot{m} / 0.3)^{8/7}$, 
or $\sim 0.2\,$yr. Given the uncertainty in the coefficients and black hole mass in 
these expressions (factors of 2--3), this is in rough agreement with the observed 
0.3--0.4$\,$yr oscillation period in J1359+4011.

\section{Conclusions}

The $15\,$GHz light curve for the flat spectrum radio quasar J1359+4011 shows a strong QPO with a period that varies between 120 and 150 days. The fluctuations in the light curve are intrinsic to the source and are not due to instrumental or observing effects. It is the only source in the OVRO catalogue of over 1200 sources to show such a strong and persistent QPO.

The physical cause of this quasi-periodic oscillation is unclear. The QPO period is of the right time-scale to be the AGN analog of LFQPOs observed at X-ray energies in stellar-mass black hole binaries. If this is the case, the precessing thick torus (that is thought by some authors to be the origin of the X-ray QPOs in black hole binaries) may be precessing the direction of the blazar jet slightly, thereby modulating the radio light curve.

Another possible explanation for the modulation of the jet luminosity is instabilities in the accretion flow. Inwardly-accreting rings caused by the Lightman-Eardley secular instability would be separated by the thermal time scale measured at the outer edge of the ``inner'' disc region. This time scale is of the right order to explain the oscillation period in J1359+4011.

\section*{Acknowledgements}
We thank Russ Keeney for his support of observations at OVRO. The OVRO 40-m program is supported in part by NASA grants 
NNX08AW31G and NNX11A043G and NSF grants AST-0808050 and AST-1109911. T.H. was supported by the Jenny and Antti Wihuri foundation. Support from MPIfR for upgrading the OVRO 40-m telescope receiver is acknowledged. We thank V. Pavlidou for useful discussions. The National Radio Astronomy Observatory is a facility of the National Science Foundation operated under cooperative agreement by Associated Universities, Inc. Part of this research was carried out at the Jet Propulsion Laboratory, California Institute of Technology, under contract with the National Aeronautics and Space Administration.

\bibliographystyle{mn2e}
\bibliography{bibliography,bibwrapper}

\end{document}